\documentclass[a4paper,fleqn,usenatbib]{mnras}

\usepackage{mathptmx}

\usepackage[T1]{fontenc}
\usepackage{ae,aecompl}

\usepackage{graphicx}	
\usepackage{amsmath}	
\usepackage{amssymb}	

\title[Nucleosynthesis of ultra-stripped SNe]{Explosive nucleosynthesis of
 ultra-stripped Type Ic supernovae: application to light trans-iron elements}

\author[T. Yoshida et al.]{
Takashi Yoshida,$^{1}$\thanks{E-mail: tyoshida@astron.s.u-tokyo.ac.jp}
Yudai Suwa,$^{2}$
Hideyuki Umeda,$^{1}$
Masaru Shibata$^{2}$
\newauthor
and Koh Takahashi$^{1,3}$
\\
$^{1}$Department of Astronomy, Graduate School of Science, University of Tokyo, Tokyo 113-0033, Japan\\
$^{2}$Center for Gravitational Physics, Yukawa Institute for Theoretical Physics, Kyoto University, 
Kyoto 606-8502, Japan\\
$^{3}$Argelander-Institute f\"ur Astronomie, Universit\"at Bonn, D-53121 Bonn, Germany
}

\date{Accepted XXX. Received YYY; in original form ZZZ}

\pubyear{2016}

\begin{document}
\label{firstpage}
\pagerange{\pageref{firstpage}--\pageref{lastpage}}
\maketitle

\begin{abstract}
We investigate the explosive nucleosynthesis during two dimensional neutrino-driven explosion 
of ultra-stripped Type Ic supernovae evolved from 1.45 and 1.5 M$_\odot$ CO stars.
These supernovae explode with the explosion energy of $\sim 10^{50}$ erg and release 
$\sim 0.1$ M$_\odot$ ejecta.
The light trans-iron elements Ga--Zr are produced in the neutrino-irradiated ejecta.
The abundance distribution of these elements has a large uncertainty because of the uncertainty
of the electron fraction of the neutrino-irradiated ejecta.
The yield of the elements will be less than 0.01 M$_\odot$.
Ultra-stripped supernova and core-collapse supernova evolved from a light CO core can be
main sources of the light trans-iron elements.
They could also produce neutron-rich nuclei $^{48}$Ca.
The ultra-stripped supernovae eject $^{56}$Ni of $\sim 0.006$--0.01 M$_\odot$.
If most of neutrino-irradiated ejecta is proton-rich, $^{56}$Ni will be produced more abundantly. 
The light curves of these supernovae indicate sub-luminous 
fast decaying explosion with the peak magnitude of about $-15$--$-16$.
Future observations of ultra-stripped supernovae could give a constraint to the event rate of 
a class of neutron star mergers.
\end{abstract}




\section{Introduction}

Mergers of binary compact objects such as neutron stars (NSs) and black holes (BHs) 
become promising sites of the r-process nucleosynthesis.
Detailed calculations of the r-process nucleosynthesis in mergers of binary compact objects succeeded in
producing heavy r-process elements with the solar abundance pattern taking account of 
general relativity and weak interactions in ejected matter 
\citep{Wanajo14,Just15,Goriely15}.
Finding the near-infrared excess in short gamma-ray burst GRB 130603B \citep{Tanvir13}
suggested that a substantial fraction of mass is likely to be ejected in the 
neutron star merger (NSM) \citep{Tanaka13,Hotokezaka13}.
On the other hand, recent chemical evolution models considering proto-galactic fragments with 
various scales and the merging time scale of $\sim 10^{7}$--$10^{8}$ yr reproduced 
the observed abundance ratio of the r-process elements 
\citep[e.g.][]{Komiya14,Tsujimoto14,Voort15,Ishimaru15,Shen15,Hirai15}.
Since the merger rate and merging time scale of binary compact objects still have uncertainties, 
it is important to constrain these values through future observations.

Ultra-stripped supernova (SN) is a candidate for generation sites of binary compact objects, 
especially binary NSs \citep{Tauris13,Tauris15, Moriya17}.
Ultra-stripped Type Ic SNe weakly explode and the destruction of the close binary system is expected to 
be suppressed \citep{Suwa15}.
The evolutionary scenarios of close binary systems of massive stars have been discussed in 
\citet{Podsiadlowski05}.
Close binary systems forming ultra-stripped Type Ic SNe would also result in the very close system of 
compact objects so that the merger of these objects takes place within cosmic age \citep{Tauris15}.
A fast-decaying faint Type Ic SN 2005ek \citep{Drout13} is considered to belong to an 
ultra-stripped SN \citep{Tauris13, Suwa15, Moriya17}.
Increasing observations of these SNe would constrain the merger rate and the merging period
of the binary compact objects.
Therefore, detailed explosion features of ultra-stripped SNe should be clarified.
It is also important to investigate the chemical composition of ultra-stripped SNe in order to 
constrain the light curve and to consider roles for the Galactic chemical evolution.

As single massive stars, the evolution of the CO cores forming a light Fe core could be similar to 
the stars slightly more massive than electron-capture (EC) SN progenitors.
These stars indicate somewhat differently evolutionary paths from more massive stars
\citep[e.g.,][]{Woosley80, Nomoto88, Umeda12, Woosley15}.
It has been found that an ECSN produces some light trans-iron elements, containing nuclei
with $Z \ge 31$, as well as neutron-rich intermediate nuclei \citep{Wanajo11}.
(The nucleosynthesis during SN explosions having a light Fe core has been done recently and 
the light trans-iron elements are also produced \citep{Wanajo17}). 
If the initial mass range of these stars is not narrow, the product composition of the SNe 
evolved from them would contribute to the Galactic chemical evolution.

In this paper, we investigate the explosive nucleosynthesis occurred in ultra-stripped Type Ic SNe
evolved from 1.45 and 1.5 M$_\odot$ CO stars.
We pursue two dimensional neutrino-radiation hydrodynamics simulations of the SN explosions 
of these stars for longer periods than \citet{Suwa15}.
Then, we calculate detailed explosive nucleosynthesis in the ejecta.
We organize this paper as follows.
In Section 2, we describe ultra-stripped Type Ic SN models evolved from the CO stars.
In Section 3, we present our nucleosynthesis calculation.
Then, we show the results of detailed explosive nucleosynthesis of the SNe.
We pay attention to $^{56}$Ni yield and the abundances of some neutron-rich isotopes and 
light trans-iron elements including the 1st peak r-process isotopes.
In Section 4, we estimate the light curves of the ultra-stripped SNe and discuss their observational 
constraints.
Then, we discuss uncertainties in the abundance of the light trans-iron elements evaluated
 in this study.
We also discuss the contribution of light trans-iron elements including light r-process isotopes 
in ultra-stripped SNe and weak SNe evolved from single massive stars 
forming a light CO core to the solar-system composition and the Galactic chemical evolution.
We conclude this study in Section 5.
In appendix A, we discuss the explosive nucleosynthesis in an ECSN.
We perform a numerical simulation for the explosion of an ECSN using the same method for the explosion 
of ultra-stripped SNe to see systematic differences in explosion models.

\section{Ultra-stripped Type Ic SN models}

\begin{table}
\centering
\caption{Properties of the ultra-stripped SN models.}
\label{tab:models}
\begin{tabular}{lcc} 
\hline
Model & CO145 & CO15 \\
\hline
Progenitor mass (M$_\odot$) & 1.45 & 1.50 \\
Mass in the computational domain (M$_\odot$) & 1.43 & 1.48 \\
Final time of the calculation (s) & 1.328 & 1.304 \\
Explosion energy (10$^{51}$ erg) & 0.170 & 0.118 \\
Ejecta mass (M$_\odot$) & 0.0980 & 0.1121 \\
Remnant baryon mass (M$_\odot$) & 1.35 & 1.39 \\
Remnant gravitational mass (M$_\odot$) & 1.24 & 1.27 \\
\hline
\end{tabular}
\end{table}

First, we perform long-term simulations of ultra-stripped SNe evolved from 
1.45 and 1.5 M$_\odot$ CO stars, i.e., CO145 and CO15 models from \citet{Suwa15}.
We follow the post bounce phase for 1.3 second.
We use a two-dimensional neutrino-radiation hydrodynamics code of  \citet{Suwa10}.
Basic features are the same as \citet{Suwa15}, but the equatorial symmetry is additionally imposed to 
save computational time.
Then, we pursue thermal evolution (i.e. temperature, density, radius, and the electron fraction) 
of the SN ejecta using a particle tracer method.
Properties of the explosions of the ultra-stripped SN models are summarized in Table \ref{tab:models}.
We obtain weak explosions with the explosion energy $\sim 10^{50}$ erg and 
small ejecta mass with $\la 0.1$ M$_\odot$.
We assume that the materials outside the computational domain of the hydrodynamics simulations 
are ejected and that they did not suffer strong explosive nucleosynthesis.

\begin{figure}
\includegraphics[width=6cm,angle=270]{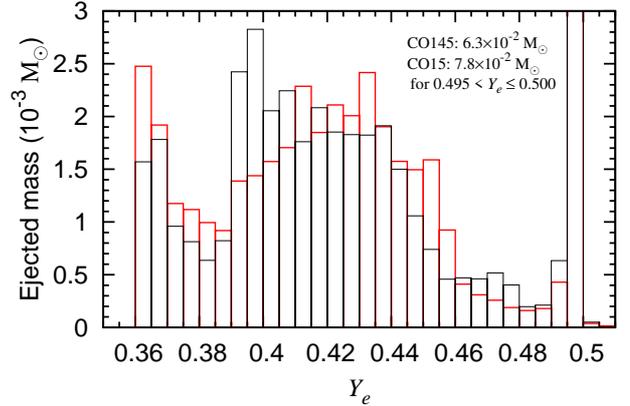}
\caption{
The ejected mass in each bin of $Y_e$ at the initial time of the nucleosynthesis calculation.
The red and black lines denote the results for CO145 and CO15 models, respectively.
The vertical axis denotes the ejected mass in units of $10^{-3}$ M$_\odot$ within each $Y_e$ bin 
of which width is $\Delta Y_e = 0.005$.
The mass in $Y_e = 0.495$--0.50 is $6.3 \times 10^{-2}$ and $7.8 \times 10^{-2}$ M$_\odot$
for CO145 and CO15 models, respectively, which is out of scale of this figure.
}
\label{fig:ye}
\end{figure}

The SN ejecta have wide distributions of the electron fraction $Y_e$.
Figure \ref{fig:ye} shows the ejected mass in each bin of $Y_e$ when the
nucleosynthesis calculation starts (see \S 3).
The electron fraction in the ejecta is distributed in $0.360 \le Y_e \le 0.508$ and $0.361 \le Y_e \le 0.503$
in CO145 and CO15 models, respectively.

Here, we define two components of the SN ejecta depending on their maximum temperature.
The shock-heated ejecta and neutrino-irradiated ejecta are defined as the ejecta whose maximum 
temperature is below and above $9 \times 10^{9}$ K, respectively.
In our models, slightly less than half of the ejecta are shock-heated ejecta.
These ejecta were in the outermost region of the progenitor.
They are ejected before they accrete onto the proto-neutron star and have not experienced electron captures.
Hence, their electron fraction is about 0.5.
The mass of the shock-heated ejecta having $Y_e$ in the range of $0.495 < Y_e \le 0.5$ is 
$6.3 \times 10^{-2}$ and $7.8 \times 10^{-2}$ M$_\odot$ for CO145 and CO15 models, respectively.
In the ejecta, the mass of the fluid components for which temperature becomes higher than 
$5 \times 10^{9}$ K is $1.1 \times 10^{-2}$ and $5.6 \times 10^{-3}$ M$_\odot$ 
for CO145 and CO15 models.
The main product of these components is $^{56}$Ni, generated through the explosive Si burning.
The other component (with the maximum temperature lower than $5 \times 10^{9}$ K) 
contains O, Ne, and other intermediate elements.

The neutrino-irradiated ejecta experience a high temperature state
and have a variety of the electron fraction depending on the neutrino irradiation.
The electron fraction of the ejecta is distributed mainly in the range between 
0.36 and 0.45.
The mass of the ejecta with $Y_e > 0.45$ is smaller than more neutron-rich ejecta.
Proton-rich materials are hardly ejected.

\begin{figure}
\includegraphics[width=6cm,angle=270]{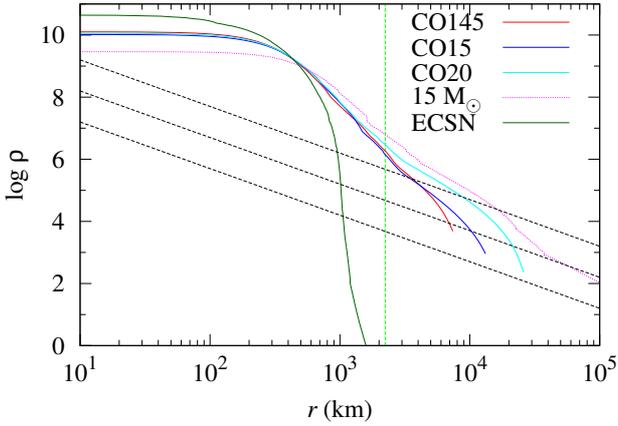}
\caption{
The density profiles of CO145 (red line), CO15 (blue line), and CO20 (cyan line) models and 
15 M$_\odot$ (pink dashed line) and ECSN progenitor (green line) models
for comparison.
The top, middle, and bottom dashed lines denote the density that deduces the accretion rate of 
0.05, $5 \times 10^{-3}$, and $5 \times 10^{-4}$ M$_\odot$ s$^{-1}$, respectively \citep{Mueller16}.
The vertical dashed line denotes 2 230 km.
See text for details.
}
\label{fig:dens}
\end{figure}

Since the $Y_e$ distributions depend on numerical methods and there have been no works 
investigating nucleosynthesis on ultra-stripped SNe, we also perform a simulation of an ECSN 
to compare our results with \citet{Wanajo11}.
The nucleosynthesis result of the ECSN is shown in Appendix A.

We note that \citet{Mueller16} discussed the condition of ECSN-like explosion using the density 
profile of SN progenitors and the critical mass accretion rate.
In his discussion, the density needs to drop to
\begin{equation}
\rho \la \frac{1}{8} \sqrt{\frac{3}{Gm}} \dot{M}_{{\rm crit}} r^{-3/2},
\label{eq:crit-dens}
\end{equation}
where $G$ is the gravitational constant, $m$ is the core mass of the progenitor, 
$\dot{M}_{{\rm crit}}$ is the critical mass accretion rate, 
$r$ is the radius.
This condition should be satisfied for a radius $r <$ 2 230 km to obtain ECSN-like explosion dynamics.
A critical mass accretion rate of $\dot{M}_{{\rm crit}} \sim 0.07$ M$_\odot$ s$^{-1}$ is expected
for ECSN-like progenitors.
Here we apply this condition to the ultra-stripped SN models CO145 and CO15.
Figure \ref{fig:dens} shows the density profiles of CO145, CO15, and CO20 \citep{Suwa15}
models and 15 M$_\odot$ \citep{Yoshida16} and ECSN models
\citep{Nomoto87} for comparison.
We also show the density profiles of Eq. (\ref{eq:crit-dens}) using $m$ = 1.4 M$_\odot$ and
$\dot{M}_{{\rm crit}}$ = 0.05, $5 \times 10^{-3}$, and $5 \times 10^{-4}$ M$_\odot$ s$^{-1}$
in accordance with \citet{Mueller16}.
The CO145 and CO15 models seem to be marginal cases for obtaining ECSN-like explosion.
The density profiles of $r \ga$ 3 000 km show steeper decline compared to s10.09 and s11.2
models in \citet{Mueller16}.

\begin{table}
\centering
\caption{1651 nuclear species adopted in the nuclear reaction network.
Isomeric state of $^{26}$Al is taken into account.}
\label{tab:net}
\begin{tabular}{lclclc} 
\hline
Element & $A$ & Element & $A$ & Element & $A$ \\
\hline
$n$ & 1 & Ca & 33--62 & Zr & 76--120 \\
H & 1--3 & Sc & 36--64 & Nb & 80--124 \\
He & 3,4,6 & Ti & 37--68 & Mo & 81--127 \\
Li & 6--9 & V & 40--71 & Tc & 84--128 \\
Be & 7,9--12 & Cr & 42--75 & Ru & 85--129 \\
B & 8,10--14 & Mn & 44--77 & Rh & 88--130 \\
C & 9--18 & Fe & 45--79 & Pd & 89--132 \\
N & 12--21 & Co & 47--81 & Ag & 92--133 \\
O & 13--22 & Ni & 48--83 & Cd & 94--135 \\
F & 17--26 & Cu & 51--86 & In & 97--136 \\
Ne & 17--29 & Zn & 52--88 & Sn & 99--137 \\
Na & 18--32 & Ga & 56--92 & Sb & 100--138 \\
Mg & 19--36 & Ge & 58--95 & Te & 114--139 \\
Al & 21--40 & As & 61--98 & I & 121--141 \\
Si & 22--43 & Se & 62--100 & Xe & 122--142 \\
P & 23--45 & Br & 66--102 & Cs & 125--143 \\
S & 24--49 & Kr & 67--107 & Ba & 126--143 \\
Cl & 28--51 & Rb & 70--110 & La & 131--143 \\
Ar & 29--55 & Sr & 71--113 & Ce & 132--143 \\
K & 32--58 & Y & 74--116 & & \\
\hline
\end{tabular}
\end{table}

\section{Explosive nucleosynthesis in the ultra-stripped Supernovae}

We calculate the explosive nucleosynthesis in the SN ejecta of CO145 and CO15 models using 
thermal history of 9968 and 8875 traced fluid particles, respectively, which have positive energy 
and positive radial velocity.
We use the nuclear reaction network consisting of 1651 nuclear species listed in Table \ref{tab:net}.
We determine the nuclear species to cover the nuclear flow in the fluid particles having
the smallest and largest $Y_e$ values using the reaction network of 5406 nuclear species 
\citep{Fujibayashi15}.

We select three initial conditions for the particles depending on the maximum temperature.
For particles of which temperature exceeds $9 \times 10^9$ K, we calculate the nucleosynthesis
from the time when the temperature decreases to $9 \times 10^9$ K.
The initial composition is set as the composition in nuclear statistical equilibrium (NSE) 
with the $Y_e$ value calculated in the hydrodynamics simulation.
For particles with the maximum temperature of (7--9) $\times 10^9$ K, we start the calculation  
from the time at the maximum temperature with the NSE initial composition.
For other particles, we calculate the nucleosynthesis from the initial time of the hydrodynamics 
simulation with the composition in the O/Ne layer.
The nucleosynthesis calculation is continued until the temperature decreases to $10^7$ K.
At the time of the termination of the hydrodynamics simulation, we determine the radial motion 
and thermal evolution assuming adiabatic expansion with the constant velocity.
We take into account the $\nu$-process in a simple manner.
The neutrino luminosity is assumed to decrease exponentially with time scale of $\tau_{\nu} = 3$ s
\citep{Woosley90}.
The total neutrino energy radiated is set to be $3 \times 10^{53}$ erg and are equipartitioned to each flavor.
The neutrino energy distribution obeys the Fermi-Dirac distribution with temperature 
$(T_{\nu_e}, T_{\bar{\nu}_e}, T_{\nu_{\mu,\tau},\bar{\nu}_{\mu,\tau}})$ = (4 MeV, 4 MeV, 6 MeV)
and zero chemical potentials \citep{Yoshida08}.
Although the adopted $\nu_{\mu, \tau}$ temperature would be larger than that obtained from recent 
spectral transport \citep[e.g.][]{Fischer10, Mueller12}, this large value scarcely affects 
the abundance distribution of the ejecta.

\begin{figure}
\includegraphics[width=6cm,angle=270]{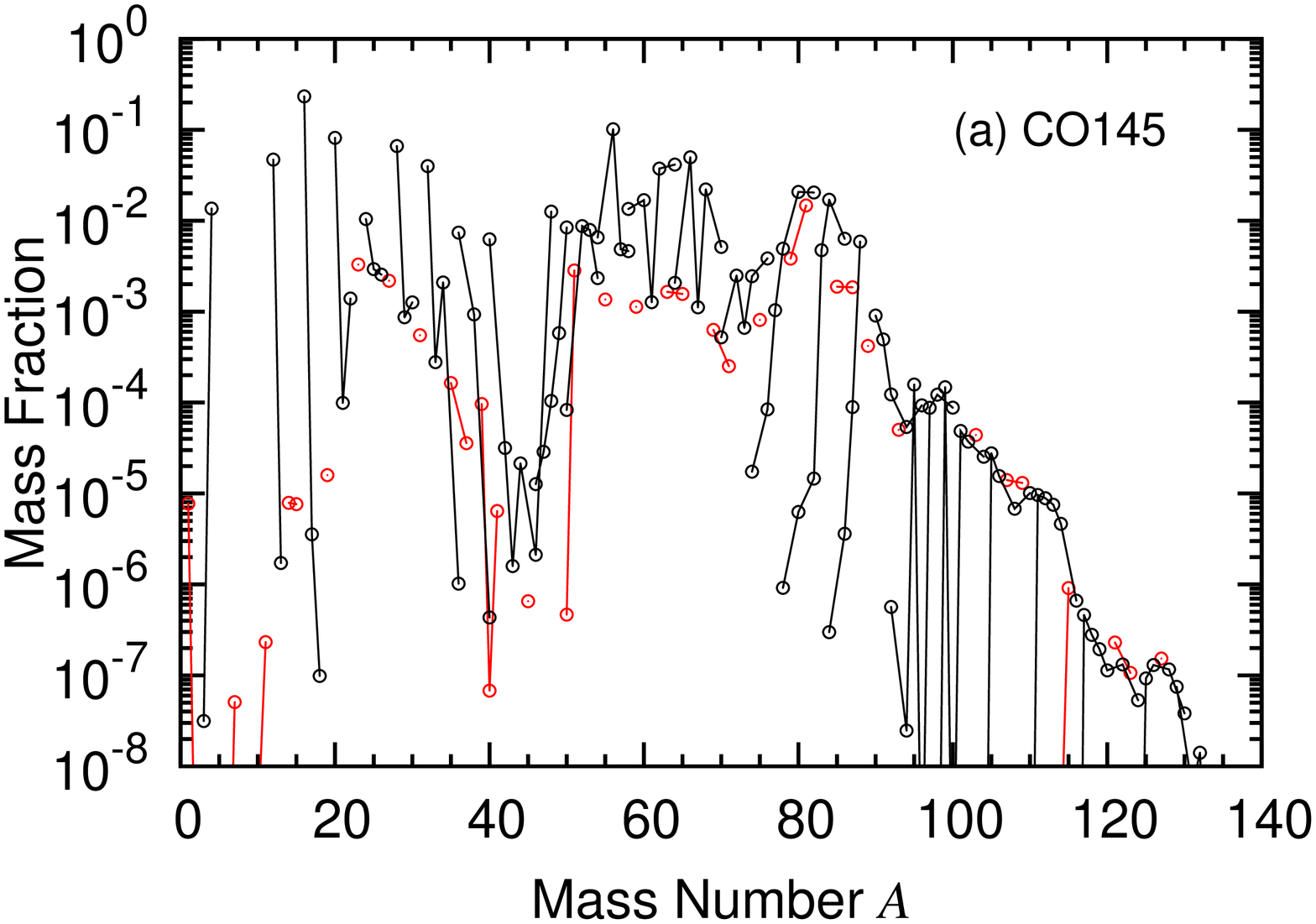}
\includegraphics[width=6cm,angle=270]{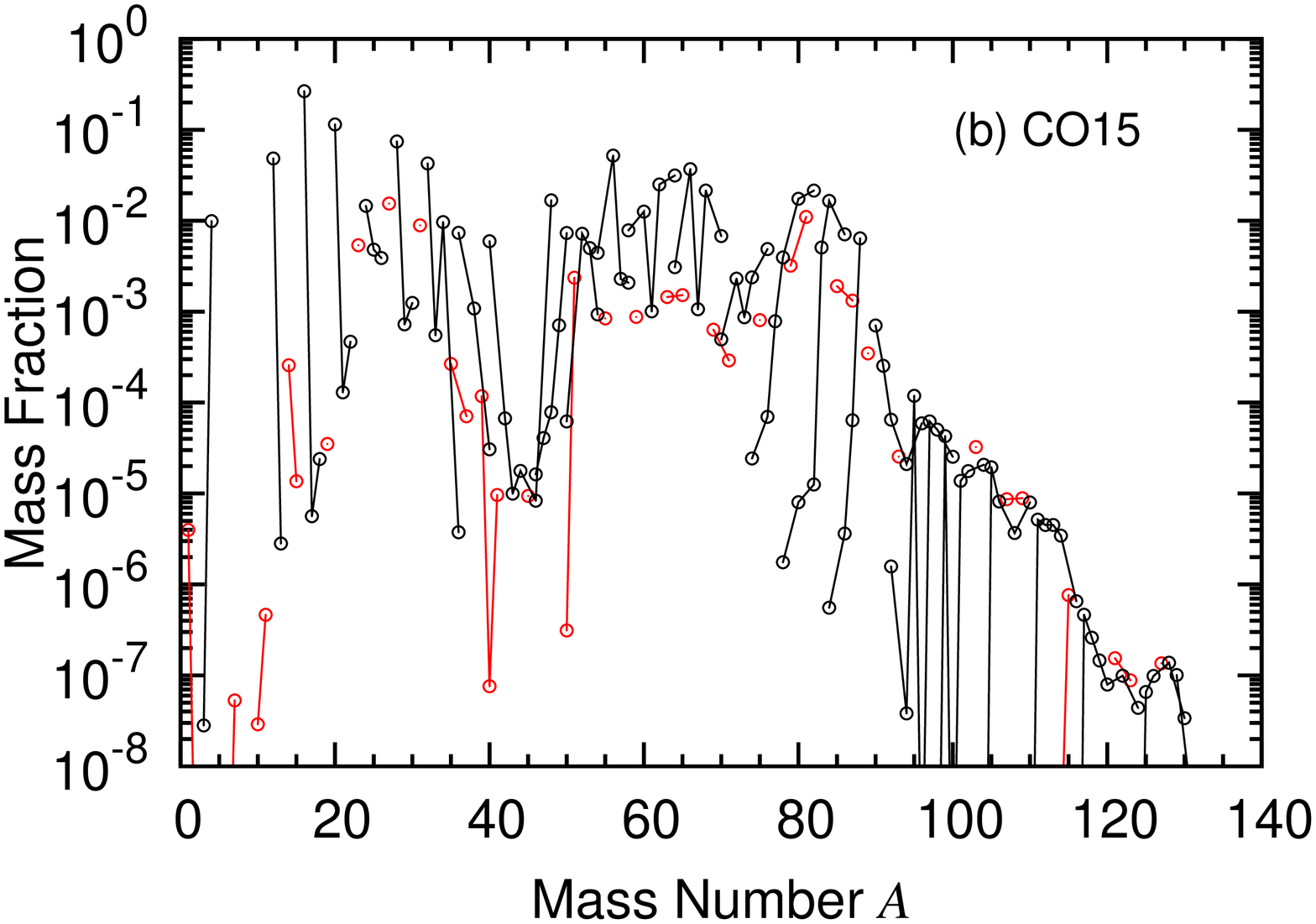}
\caption{
Mass fraction distribution of isotopes in the ejecta of ultra-stripped Type Ic SNe.
Panels (a) and (b) denote CO145 and CO15 models, respectively.
The red and black lines correspond to odd-$Z$ and even-$Z$ isotopes.
}
\label{fig:massfrac}
\end{figure}

Figure \ref{fig:massfrac} shows the mass fraction distribution of isotopes in the SN ejecta
of CO145 and CO15 models.
Yields of some elements and isotopes ejected in CO145 and CO15 models are listed in Table \ref{tab:yield}.
General features are not different between CO145 and CO15 models.
Elements with the mass of $A \la 90$ are broadly produced with the maximum mass fraction of 0.3.
Elements with $90 \la A \la 130$ are also produced but their mass fractions decrease with 
mass number.

Most of C, O, and intermediate nuclei with $A \la 40$ are mainly unburned or synthesized
through explosive O burning.
Light iron peak elements, Ti, V, and Cr, are produced in neutron rich ($Y_e \la 0.40$) materials.
Mn and Fe are produced through explosive Si burning.
The $^{56}$Ni yield is $9.73 \times 10^{-3}$ and $5.72 \times 10^{-3}$ M$_\odot$
in CO145 and CO15 models, respectively.
These values are smaller than the expectation in \citet{Suwa15}.
This is because some of the materials that experienced a state of temperature higher than 
$5 \times 10^9$ K are neutrino-irradiated ejecta.
They have become neutron rich and are synthesized to be lighter and heavier elements.
Heavy neutron-rich isotopes of $A \sim 60 - 90$ are also produced in the neutrino-irradiated ejecta
containing neutron rich materials.
The 1st peak r-process isotopes such as $^{72-74,76}$Ge, $^{75}$As, $^{77,78,80,82}$Se, 
$^{81}$Br, $^{83,84,86}$Kr, $^{85,87}$Rb, $^{88}$Sr \citep[Table 1 in][]{Sneden08} are included.
They are produced in nuclear quasi-equilibrium \citep{Meyer98, Wanajo11}.
On the other hand, most of $^{56}$Ni is produced in the shock-heated ejecta.

\begin{figure}
\includegraphics[width=6cm,angle=270]{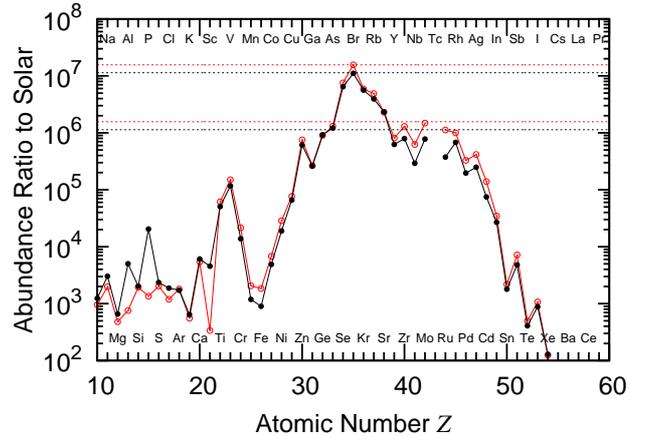}
\caption{
Abundance ratios of elements in the ejecta of ultra-stripped Type Ic SN to the solar abundance.
The red and black lines denote the ratios of CO145 and CO15 models, respectively.
The dashed lines denote the maximum ratios and the ratios of the 10\% of the maximum ratios.
}
\label{fig:abundratio_z}
\end{figure}

We consider the contribution to the solar-system composition.
Figure \ref{fig:abundratio_z} shows the elemental abundance ratios to the solar-system composition.
The elemental abundance ratio means that the ratio of the abundance in the SN ejecta to the normalized 
solar abundance in \citet{Asplund09} for each element.
Elements As--Sr indicate large abundance ratios, 
i.e., more than 10\% of the largest abundance ratio.
The element of the largest abundance ratio is Br.
The ratios of heavier elements Y--Rh are smaller than the above elements but the ratios are still 
larger than other elements.
Se, Br, and Kr are mainly produced in the ejecta of $Y_e \sim 0.36$--0.43.
On the other hand, Sr, Y, and Zr are produced in $Y_e \sim 0.36$--0.38 as well as 0.42--0.46.
Since the ejecta mass of $Y_e \ga 0.44$ materials is small, the contribution of Sr--Zr is smaller than
Se--Kr.
The light trans-iron elements produced in ultra-stripped SNe may contribute
to the solar system abundance and the Galactic chemical evolution.
More details of the $Y_e$ dependence of the synthesis of trans-iron elements
will be discussed in \S 4.2.

\begin{table*}
\centering
\caption{Yields of some elements and isotopes in units of $10^{-3}$ M$_\odot$.}
\label{tab:yield}
\begin{tabular}{lcclcclcclcc} 
\hline
Element & CO145 & CO15 & Element & CO145 & CO15 & 
Isotope & CO145 & CO15 & Isotope & CO145 & CO15 \\
\hline
C & 4.60 & 5.41 & Si & 6.71 & 8.56 & 
$^{48}$Ca & 1.24 & 1.88 & $^{53}$Mn & 0.00200 & 0.00203 \\
O & 22.9 & 29.7 & S & 4.14 & 5.91 & 
$^{50}$Ti & 0.825 & 0.821 & $^{56}$Ni & 9.73 & 5.72 \\
Ne & 8.12 & 12.9 & Ca & 4.14 & 5.91 & 
$^{26}$Al & 0.214 & 0.211 & $^{60}$Fe & 0.528 & 0.343 \\
Mg & 1.56 & 2.61 & $Z \ge 31$ & 11.6 & 12.4 & 
$^{41}$Ca & 0.000593 & 0.000798 \\
\hline
\end{tabular}
\end{table*}

\begin{figure}
\includegraphics[width=6cm,angle=270]{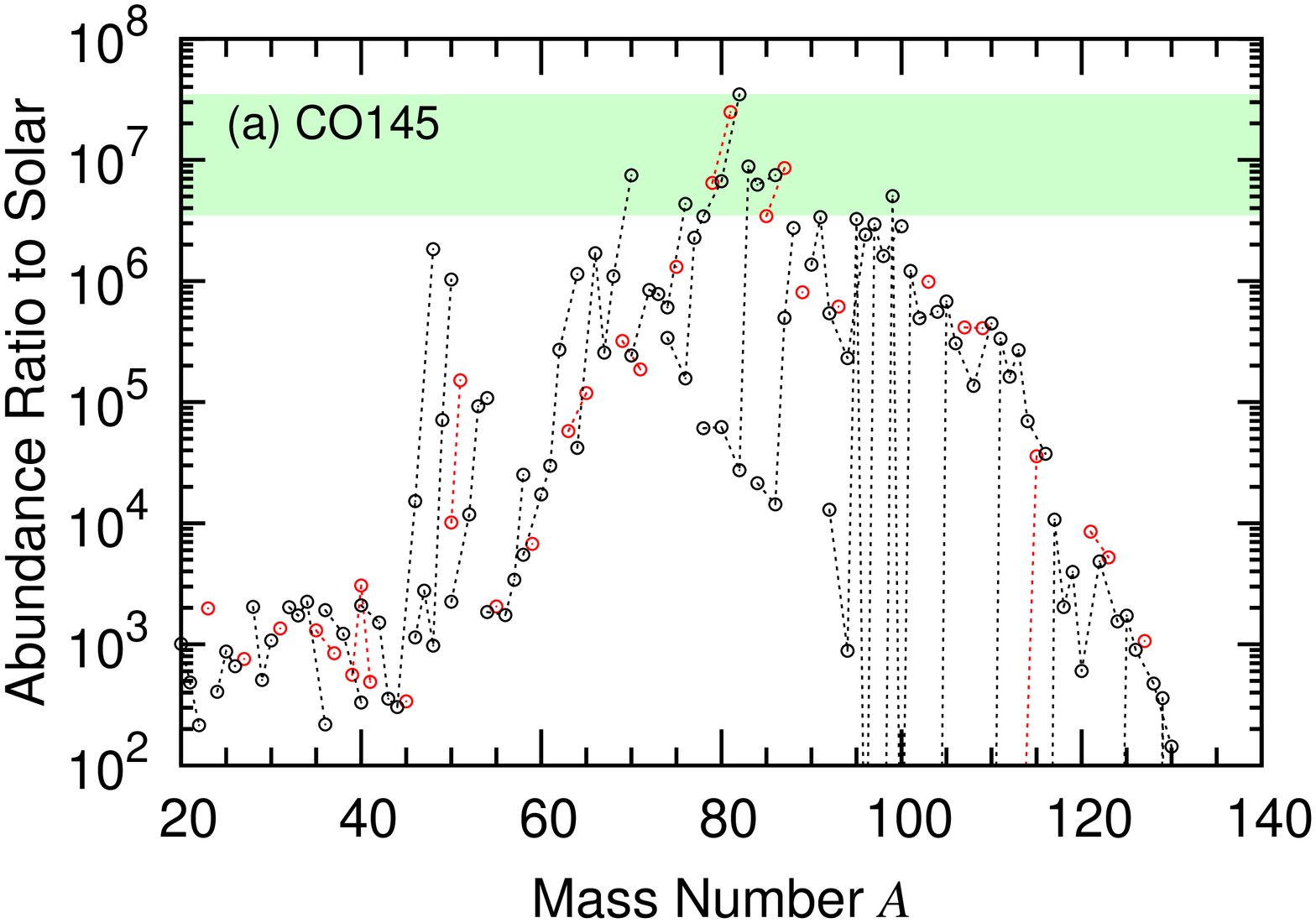}
\includegraphics[width=6cm,angle=270]{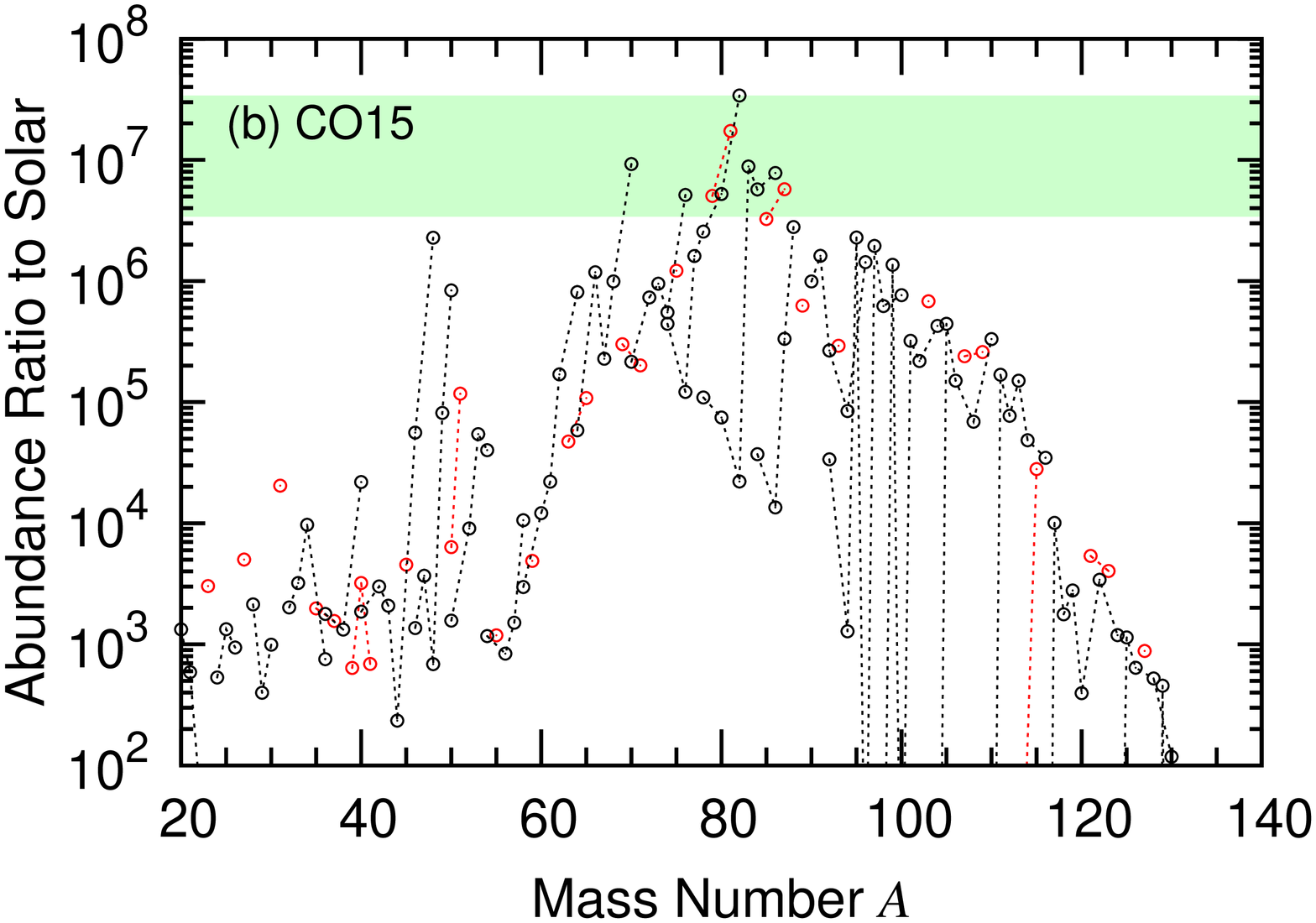}
\caption{
Abundance ratios of isotopes in the ejecta of CO145 (panel (a)) and CO15 (panel (b)) models 
to the solar abundance.
The red and black lines correspond to odd-$Z$ and even-$Z$ nuclei.
The green hatched region denotes the range of the abundance ratio more than 10\% of the maximum ratio.
}
\label{fig:abundratio_a}
\end{figure}

Figure \ref{fig:abundratio_a} shows the isotopic abundance ratios to the solar-system composition
in CO145 and CO15 models.
The isotopic abundances in the solar composition are adopted from \citet{Lodders09}.
Yields of some isotopes are also listed in Table \ref{tab:yield}.
The isotopes of $A \sim 70$--100 indicate large abundance ratios.
We also see large ratios of $^{48}$Ca.
The obtained ratio of $^{48}$Ca is similar to ECSN.
Although it has been shown that $^{48}$Ca is produced in low-entropy neutron-rich 
($Y_e \sim 0.42$) expansions \citep{Meyer96}, the origin of $^{48}$Ca has not been clarified.
A possibility of the production in neutron-rich ejecta of SNe Ia was proposed in \citet{Woosley97}.
ECSN has been pointed out as a possible site of $^{48}$Ca \citep{Wanajo13}.
Ultra-stripped SNe and core-collapse SNe evolved from light CO cores are possible sites
of $^{48}$Ca.

\section{Discussion}

\subsection{Light curves of ultra-stripped SNe}

We derive light-curve models of ultra-stripped SNe of CO145 and CO15 models 
using the analytic solution shown in \citet{Arnett82}.
The energy generation rates of $^{56}$Ni and $^{56}$Co are adopted from \citet{Nadyozhin94}.
We use the deposition factor of $^{56}$Co as
$D_{\rm Co} = 0.968 D(\tau_\gamma) + 0.032 D(355\tau_\gamma)$, 
where $\tau_\gamma$ is the optical depth of $\gamma$-rays, to take into account 
both the $\gamma$-rays energy release and the positron kinetic energy release 
\citep{Nadyozhin94, Colgate97}.
We adopt $\kappa = 0.1$ cm$^2$ g$^{-1}$ for the opacity of the SN ejecta.
We include the effect of gamma-ray leakage approximately using the deposition function as 
shown in \citet{Arnett82} \citep[see also][]{Colgate80}.
The corresponding gamma-ray opacity is $\kappa_\gamma = 0.03$ cm$^{2}$ g$^{-1}$.
We consider the energy release by the radioactive decays of intermediate and heavy elements.
We are not sure about the fractions of the energy deposition by gamma-rays and positrons from their 
radioactive decays.
Hence, we assume the energy deposition fractions by gamma-rays and positrons as 0.5
for simplicity.

\begin{figure}
\begin{center}
\includegraphics[width=6cm,angle=270]{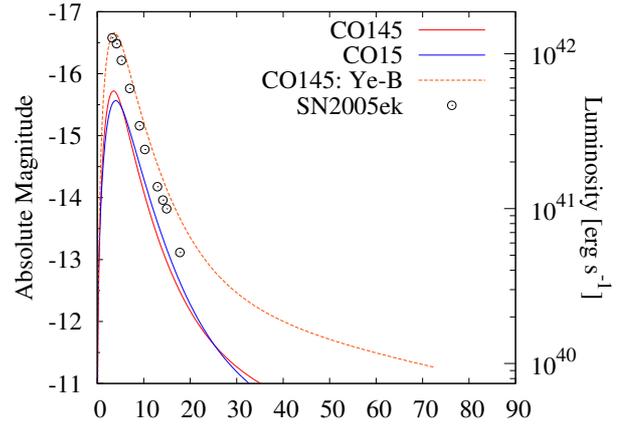}
\end{center}
\caption{
Light curves of the ultra-stripped Type Ic SNe.
The red and blue curves denote CO145 and CO15 models, respectively.
The circles denote the light curve of SN 2005ek.
The orange dashed line denotes CO145 Ye-B model (see \S 4.2).
}
\label{fig:lc}
\end{figure}

Figure \ref{fig:lc} shows the light curves of these SNe.
For comparison, we also show the light curve of a fast decaying Type Ic SN 2005ek, 
of which ejecta mass and $^{56}$Ni mass are estimated as 0.1 $M_\odot$ and 0.03 M$_\odot$
by \citet{Drout13}.
The peak absolute magnitude of our models is $-15.5$--$-16$.
We also vary the yield ranges for $^{56}$Ni and other radioactive isotopes between 
twice and a half of values numerically obtained.
Then, the range of the peak absolute magnitude varies in the range $-14.8$--$-16.5$.
When the yields of $^{56}$Ni and other radioactive isotopes are twice in CO145 model, 
the light curve reproduces that of SN 2005ek.

The energy release of the radioactive decays of iron-peak elements other than $^{56}$Ni and 
$^{56}$Co partly contributes to the optical emission. 
In CO145 model, the radioactive decays of these elements have a contribution for one day 
after the collapse.
The fraction of the luminosity from these elements is 27 \% at the peak luminosity.
On the other hand, the radioactive decays from these elements partly contribute
for four days in CO15 model.
If the contribution from these elements is ignored, the peak luminosity becomes about a half.
The main energy source other than $^{56}$Ni and $^{56}$Co is $^{57}$Ni and $^{66}$Cu for
CO145 and CO15 models, respectively.
The decay time of the luminosity from these elements is about four and eight days 
in CO145 and CO15 models, respectively.
The difference of the contribution from these elements is mainly due to the difference in the $^{56}$Ni yield.

Recently, a variety of fast decaying SNe have been found in survey programs for transient objects.
Sub-luminous SNe have also been observed in Types Ia and Ib/c SNe \citep[e.g.][]{Foley13, Drout14}.
Some of sub-luminous fast decaying SNe could be ultra-stripped SNe.
These observed SNe showed spectral features different from normal Types Ia and Ib/c SNe.
The ejecta of the ultra-stripped SNe in our models indicate a larger abundance ratio of intermediate
elements to oxygen compared to more massive CO cores.
These compositional differences could give distinctive spectra features.
Identification of ultra-stripped SNe from Type I SNe is important for the evaluation of 
the ultra-stripped SN rate.
Future observations of ultra-stripped SNe could constrain the rates of 
ultra-stripped SNe.

We note, as pointed out in \citet{Suwa15}, that it is safe to consider that our results give a lower limit 
of the explosion energy of an ultra-stripped SN.
In the case of stronger explosion of the ultra-stripped SN, the ejected $^{56}$Ni mass could be larger.
If so, ultra-stripped SNe could be observed as fast decaying SNe like Type Ic SN 2005ek.
We also note that the $^{56}$Ni mass would be underestimated because of the missing 
proton-rich component in the neutrino-irradiated ejecta.
This will be discussed in \S 4.2.

\begin{figure}
\begin{center}
\includegraphics[width=6cm,angle=270]{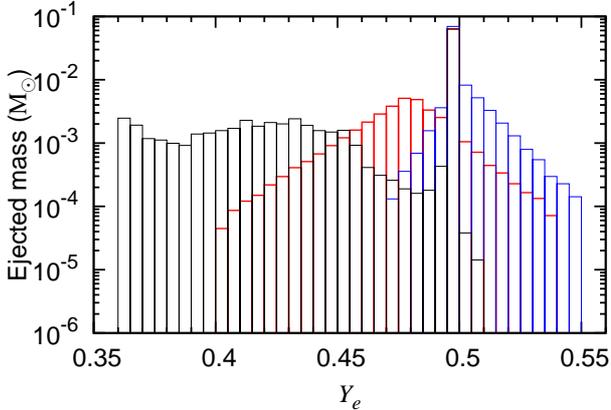}
\end{center}
\caption{
Same as Fig. \ref{fig:ye}, but for the modified $Y_e$ cases of CO145 model.
Red, blue, and black lines denote CO145-Ye-W, CO145-Ye-B, and unmodified CO145 models,
respectively.
}
\label{fig:mass-yewb}
\end{figure}

\subsection{Uncertainties of the yield of light trans-iron elements in ultra-stripped SN models}

We obtained light trans-iron elements in the ultra-stripped SN models.
However, the production efficiency of the elements depends on the $Y_e$ distribution of the SN ejecta, 
which also depends on detailed treatment of neutrino transport.
Indeed, \citet{Mueller16} showed that an approximate treatment of neutrino transport introduces 
a broader $Y_e$ distribution compared to a more stringent model including sophisticated microphysics.
On the other hand, an update of code can even lead to a smaller $Y_e$ distribution.
For instance, an ECSN simulation performed by Garching group with an updated code showed 
a smaller minimum value of $Y_e$ (0.34) than the previous result \citep[0.404;][]{Wanajo11}
(Janka 2016, private communication\footnote{see also 
http://www2.yukawa.kyoto-u.ac.jp/\texttt{\char`\~}npcsm\\/conference/slides/01Tue/Janka.pdf}).
Therefore, it is turned out that $Y_e$ distribution is rather sensitive to the detailed treatment of
microphysics, including neutrino transfer method.
To assess the uncertainties originated from the difference of codes, we perform numerical 
simulations of an ECSN with our code used in the simulations of the ultra-stripped SNe and carry out a 
simulation of the explosive nucleosynthesis. 
We compare the ECSN yield with the result in \citet{Wanajo11} and find that  
more neutron-rich materials are ejected in our model (see Appendix A).

To study the uncertainties of the yield of trans-iron elements more systematically, 
we here additionally consider
two different ways of the modification of the $Y_e$ distributions of CO145 model.
As the first way, we consider three cases of the $Y_e$ distributions by increasing the minimum 
$Y_e$ value.
We refer to the three models as Ye040, Ye042, and Ye044 models.
Ye$XXX$ indicates that the minimum $Y_e$ value in the corresponding model, $Y_{e, {\rm min}}$, 
is $X.XX$.
We modify the $Y_e$ values of the tracer particles by 
\begin{equation}
Y_{e, {\rm mod}} = \frac{0.50 - Y_{e, {\rm min}}}{0.14} (Y_e -0.36) + Y_{e, {\rm min}}
\end{equation}
for
$Y_e \le 0.5$ in each model.

As the second way, we consider two cases of the $Y_e$ distributions based on 
\citet{Wanajo11} and \citet{Buras06}, in which the $Y_e$ distributions of ECSN and 
15 M$_\odot$ CCSN models are shown respectively.
In the ECSN model, the $Y_e$ value of the ejecta distributes between 0.404 and 0.54 
\citep[Fig. 2 in][]{Wanajo11}.
The peak of the ejecta mass is at $Y_e \sim 0.48$.
In the 15 M$_\odot$ CCSN model, the $Y_e$ value of the ejecta distributes between 0.470 and 0.555
\citep[Fig. 41 in][]{Buras06}.
The ejecta mass in each $Y_e$-bin with $\Delta Y_e = 0.005$ varies in the range of 
$10^{-4}$--$0.01$ M$_\odot$ and the peak is located at $Y_e \sim 0.50$.
We refer to the former and the latter models as Ye-W and Ye-B models, respectively.
We construct two modified $Y_e$ models having similar $Y_e$ distribution of these two models
as follows.
First, we pick up tracer particles in the neutrino-irradiated ejecta and calculate the mass fraction to 
the total mass of the neutrino-irradiated ejecta for each particle.
Next, we set the particles in the increment order of $Y_e$.
We calculate the cumulative mass fraction $c_M$ as a function of $Y_e$ for each particle.
Then, we set the modified $Y_e$ values using the following equations
\begin{equation}
Y_{e,{\rm mod}} = \left\{
\begin{array}{l}
  0.48 + 0.04 \log_{10} (1.73 c_M +0.01) \quad \mbox{for $c_M \le 4/7$} \\
  0.48 - 0.06 \log_{10} (2.31 (1-c_M) +0.01) \quad \mbox{for $c_M \ge 4/7$}
\end{array}
\right.
\end{equation}
for Ye-W model and
\begin{equation}
Y_{e,{\rm mod}} = \left\{
\begin{array}{l}
  0.5 + 0.015 \log_{10} (2.64 c_M +0.01) \quad \mbox{for $c_M \le 0.375$} \\
  0.5 - 0.025 \log_{10} (1.58 (1-c_M) +0.01) \quad \mbox{for $c_M \ge 0.375$}
\end{array}
\right.
\end{equation}
for Ye-B model.
The obtained $Y_e$ distribution is shown in Figure \ref{fig:mass-yewb}.
The mass fraction of the particles of $Y_e > 0.5$ in the neutrino-irradiated ejecta is 0.09 and 0.63 
for Ye-W and Ye-B models, respectively.
We do not modify the $Y_e$ values of the shock-heated ejecta.

\begin{figure}
\begin{center}
\includegraphics[width=6cm,angle=270]{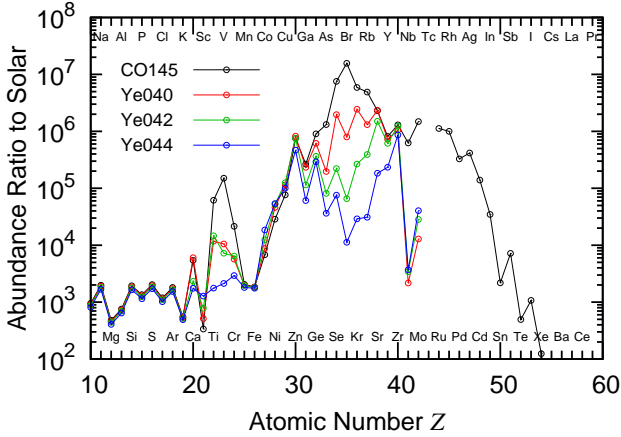}
\end{center}
\caption{
Abundance ratios of elements in the ejecta of ultra-stripped SN models to the solar abundance.
The black, red, green, and blue lines denote the ratios of CO145, Ye040, Ye042, and Ye044
models, respectively.
}
\label{fig:yedep-el}
\end{figure}

First, we present the result of Ye040, Ye042, and Ye044 models.
Figure \ref{fig:yedep-el} shows the elemental abundance ratios to the solar-system composition.
Light trans-iron elements are produced in the three models.
However, the dependence of the abundance ratios on the $Y_e$ distribution has a variety in the elements.
The abundance ratios of Ge and Zr depend scarcely on the $Y_e$ distribution in these $Y_e$ modifications.
By contrast, the abundance ratios of the elements around the 1st r-peak such as Se and Br 
depend strongly on.
In Ye040 model, we find a moderate variation in the abundance ratios among Ga--Zr.
The Ye044 model indicates the decrease in the abundance ratios in Ga and As--Rb.
We also see large $Y_e$ dependence in Ti--Cr.

\begin{figure}
\begin{center}
\includegraphics[width=6cm,angle=270]{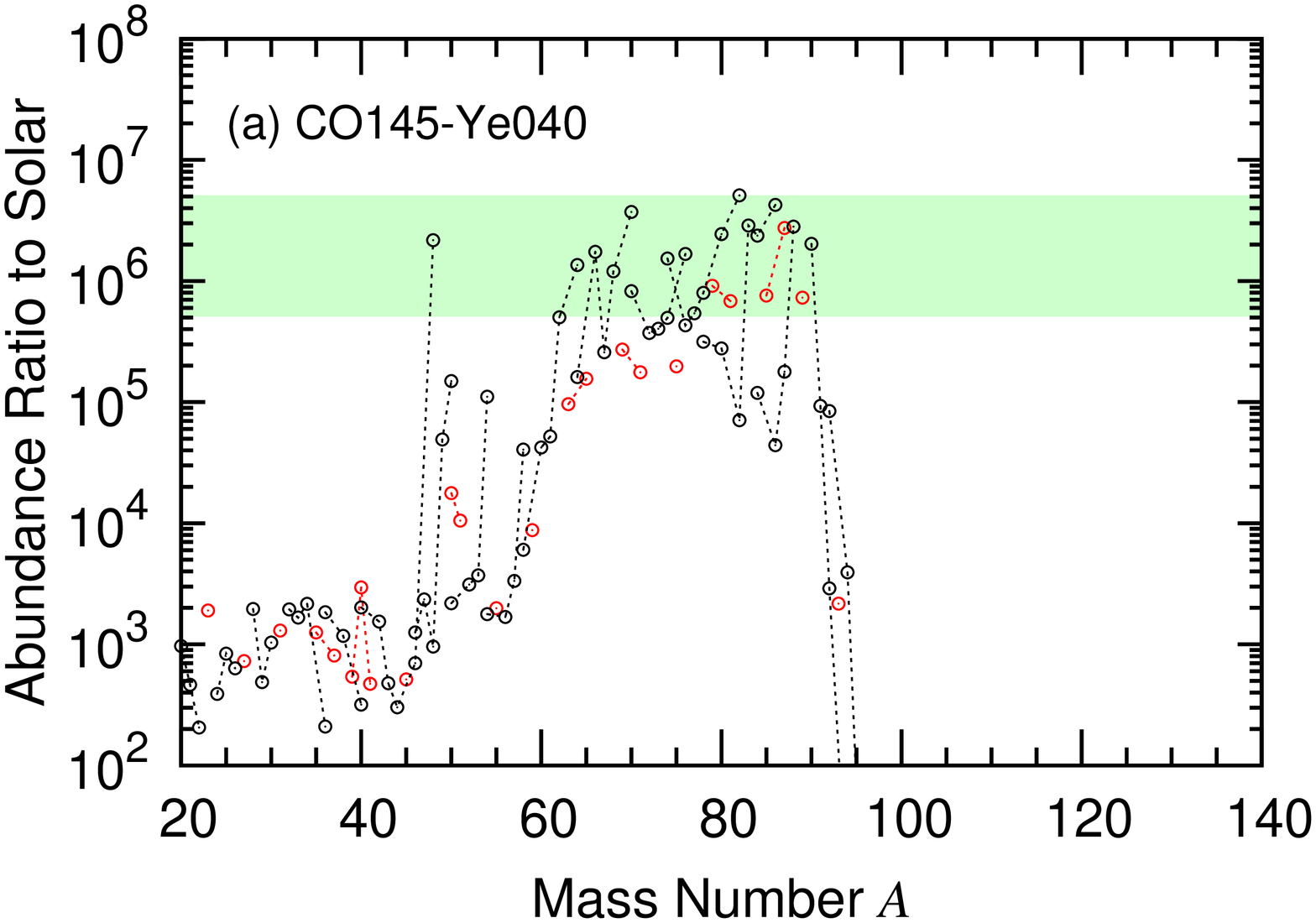}
\includegraphics[width=6cm,angle=270]{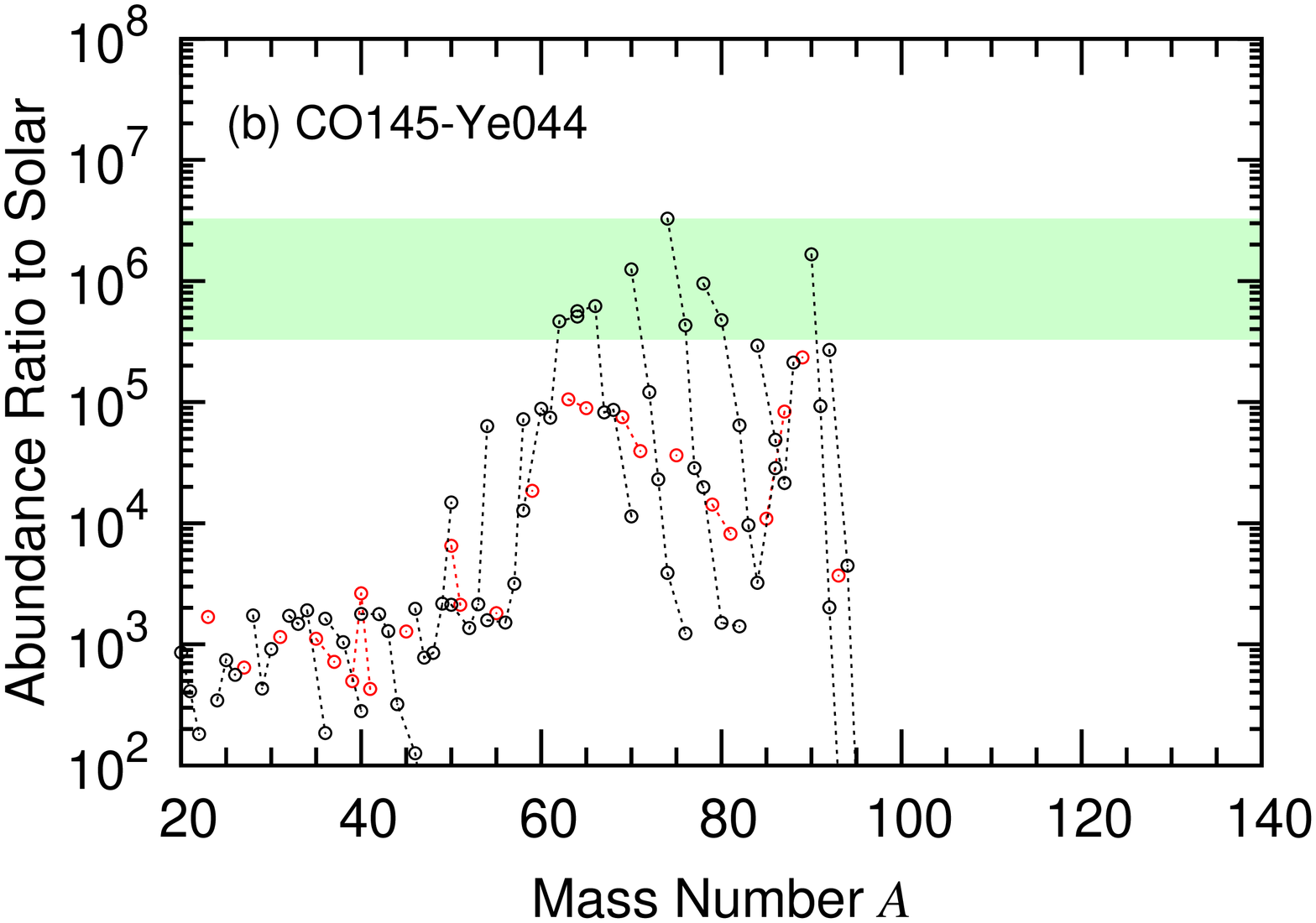}
\end{center}
\caption{
Same as Fig. \ref{fig:abundratio_a} but for Ye040 model (panel (a)) and Ye044 model (panel (b)).
}
\label{fig:yedep-iso}
\end{figure}

Figure \ref{fig:yedep-iso} shows the isotopic abundance ratios to the solar-system composition 
for Ye040 and Ye044.
We find a clear difference in the isotopic abundance ratios in the two models.
The distribution of Ye040 model indicates large abundance ratios for neutron-rich nuclei
for most of Fe-peak elements and trans-iron elements up to Sr.
We also find the decrease in the abundance ratios of neutron-rich isotopes for Zr and Mo.
In Ye044 model, on the other hand, a large abundance ratio of neutron-rich nuclei is limited 
in iron-peak elements up to Ni.
For the elements heavier than Zn except for Sr, p-nuclei rather than neutron-rich nuclei indicate 
large abundance ratios.
The production of the light p-nuclei in such slight neutron-rich materials has been found 
in \citet{Hoffman96}.
Thus, the isotopic abundance distribution also depends on the $Y_e$ distribution of the SN ejecta.

\begin{figure}
\begin{center}
\includegraphics[width=6cm,angle=270]{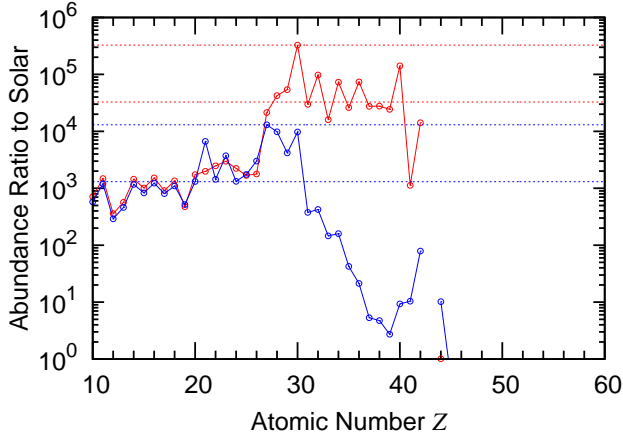}
\end{center}
\caption{
Same as Fig. \ref{fig:yedep-el}, but for Ye-W (red line) and Ye-B (blue line) models.
}
\label{fig:el-yewb}
\end{figure}

Next, we present the abundance distributions of Ye-W and Ye-B models.
Figure \ref{fig:el-yewb} shows the elemental abundance ratios to the solar-system composition 
for the two models.
The isotopic abundance ratios to the solar-system composition are shown 
in Fig. \ref{fig:abundratio_a-yewb} in Appendix B.
The large abundance ratios of the light trans-iron elements are obtained 
in Ye-W model.
Zr shows the largest abundance ratio among the elements.
On the other hand, the large abundance ratio is shown in iron-peak elements with $Z \sim 22$--30 
in Ye-B model.
The production of the light trans-iron elements in Ye-B model does not contribute 
to the solar-system composition.

We note that the $^{56}$Ni yield is $2.9 \times 10^{-2}$ M$_\odot$ in Ye-B model.
This yield is three time as large as in the original CO145 model.
We draw the light curve of Ye-B model in Fig. \ref{fig:lc}.
The maximum absolute magnitude is close to SN 2005ek.
We also note that the large amount of Sc is produced in Ye-B model.
Most of Sc is produced in the proton-rich neutrino-irradiated ejecta.
This production process of Sc has been obtained in \citet{Pruet05} and \citet{Frohlich06}.

In order to understand the obtained abundance features, we investigate the $Y_e$ dependence 
of elemental mass fractions and abundant isotopes for even elements in the ejecta of CO145 model.
The Ge mass fraction distribution on $Y_e$ has two peaks at $Y_e \sim 0.39 - 0.40$ and 
$Y_e \sim 0.46$.
The former peak is about five times larger than the latter.
Sr and Zr have also two peaks in their mass fraction distribution on $Y_e$.
The one is at $Y_e \sim 0.36$.
The other is in $Y_e \sim 0.43 - 0.45$ and $\sim 0.46$ for Sr and Zr, respectively.
We find that the mass fraction distributions of Ga, Ge, and Sr--Zr have similar $Y_e$ dependence and
that these elements are produced in the ejecta of $Y_e \ga 0.44$.
Thus, the yields of these elements in the SN ejecta depend scarcely on the uncertainty of the 
$Y_e$ distribution.

On the other hand, the mass fractions of Se and Kr are larger in the range of 
$Y_e \la 0.41$ and 0.43, respectively.
The mass fraction distribution of Mo has a peak at $Y_e \sim 0.36$.
Although Mo is also produced in the ejecta of $Y_e \sim 0.47$, the mass fraction in the ejecta is 
smaller than the neutron-rich ejecta by more than two orders of magnitude.
Thus, Se--Rb, Nb, and Mo have large dependence on the $Y_e$ distribution.

The abundant isotopes also depend on $Y_e$.
In the ejecta of $0.36 \le Y_e \le 0.40$, the isotopes of which mass fraction $X$ is larger than 0.01 are
$^{72}$Ge, $^{76}$Ge, $^{78,80,82}$Se, $^{79,81}$Br, and $^{83,84}$Kr.
The mass fractions of the isotopes $^{74}$Se, $^{75}$As, $^{77}$Se, and $^{85,87}$Rb exceed 
$5 \times 10^{-3}$.
In the less neutron-rich ejecta of $0.40 < Y_e \le 0.44$, the abundant isotopes ($X > 0.01$) are
$^{78,80,82}$Se, $^{84,86}$Kr, and $^{88}$Sr.
The secondary abundant isotopes ($X > 5 \times 10^{-3}$) are $^{74,76}$Ge, $^{83}$Kr, and $^{87}$Rb.
In the ejecta of $0.44 < Y_e \le 0.48$, the isotopes with $X > 5 \times 10^{-3}$ are
$^{70,72}$Ge, $^{88}$Sr, and $^{90}$Zr.
The isotopes in the 1st r-process peak, 
$^{76}$Ge, $^{75}$As, $^{80}$Se, $^{82}$Se, and $^{81}$Br, 
are mainly produced in the ejecta of $Y_e \sim$ 0.36--0.42.
A similar dependence of the abundantly produced isotopes on $Y_e$ has been shown
 in the nucleosynthesis of an ECSN \citep{Wanajo11}.

The yield of the light trans-iron elements also depends on the $Y_e$ distribution.
The yields in Ye040, Ye042, and Ye044 models are $4.2 \times 10^{-3}$, 
$1.4 \times 10^{-3}$, and $6.5 \times 10^{-4}$ M$_\odot$, respectively.
They decrease with increasing the minimum $Y_e$ value in the
$Y_e$ distributions.
The yields of the light trans-iron elements in Ye-W and Ye-B models are 
$3.0 \times 10^{-4}$ M$_\odot$ and $1.1 \times 10^{-6}$ M$_\odot$, respectively.
The former model has the minimum $Y_e$ value of 0.40 but the amount of $Y_e \le 0.44$ ejecta
is only 6 \% of the total ejecta.
The latter model shows the largest minimum $Y_e$ value among the modified $Y_e$ models.
This is due to the $Y_e$ dependence of the yield.
We investigate the mass fractions of the light trans-iron elements with the relation to the $Y_e$ value
in the ejecta.
In the ejecta with $Y_e \la 0.40$, the mass fraction is roughly 0.7.
It decreases with increasing $Y_e$ for the larger $Y_e$ ejecta.
It is less than 0.5, 0.2, and 0.1 for the ejecta of $Y_e \ga 0.41$, 0.43, and 0.45, respectively 
in CO145 model.
The above dependence is commonly seen in the modified $Y_e$ models.
We also note that the total mass of the neutrino-irradiated ejecta depends on the treatment of 
neutrino transport of explosion models.
This effect will increase the uncertainty in the yield of the light trans-iron elements.

We consider that the trans-iron elements up to Zr are produced in ultra-stripped SNe.
However, their elemental and isotopic abundances and the total yield depend on the $Y_e$ distribution
and the minimum $Y_e$ value in the SN ejecta.
In this study, the yield of light trans-iron elements is between $6 \times 10^{-4}$ and 0.01 M$_\odot$.

The $^{48}$Ca yield also depends on the uncertainty of the $Y_e$ distribution in the 
SN ejecta.
The $^{48}$Ca yield is $1.5 \times 10^{-3}$, $1.8 \times 10^{-4}$, and $1.0 \times 10^{-8}$ M$_\odot$
in Ye040, Ye042, and Ye044 models, respectively.
We also obtain the $^{48}$Ca yield of $8.1 \times 10^{-5}$ M$_\odot$ in Ye-W model.
It has been found that $^{48}$Ca is produced in low-entropy neutron-rich ($Y_e \sim 0.42$) materials
\citep{Meyer96}.
The very small $^{48}$Ca yield is due to the lack of the production site.
Ultra-stripped SNe would be a production site of $^{48}$Ca.
Reducing the uncertainty in the $Y_e$ distribution will also constrain the $^{48}$Ca yield
in ultra-stripped SNe.

\subsection{Contribution of the 1st peak r-process isotopes to the Galactic chemical evolution}

We obtained the light trans-iron elemental yield of $\sim 6 \times 10^{-4}$ to 0.01 M$_\odot$ 
containing the 1st peak r-process isotopes for CO145 and CO15 models taking into account 
the assumed uncertainty in the electron fraction distributions.
Here, we discuss the contributions of ultra-stripped SNe to the solar-system composition and
the Galactic chemical evolution for the r-process isotopes.
We consider  $^{72-74,76}$Ge, $^{75}$As, $^{77,78,80,82}$Se, $^{81}$Br, $^{83,84,86}$Kr, 
$^{85,87}$Rb, $^{88}$Sr as the 1st peak r-process isotopes \citep{Sneden08}.

Possible sites of heavy r-process elements were investigated using the abundance of 
short-lived isotope $^{244}$Pu in interstellar particles found in earth's deep sea floor \citep{Hotokezaka15}.
They found that the production of more than 0.001 M$_\odot$ r-process elements 
by NSMs with the NSM rate smaller than 90 Myr$^{-1}$ can explain the
production of the r-process elements in the Galaxy.
The near-infrared excess of GRB 130603B indicated the ejecta mass of the NSM should be larger 
than 0.02 M$_\odot$ \citep{Hotokezaka13}.
Recent numerical simulations of NSMs \citep{Wanajo14,Sekiguchi15,Sekiguchi16} and 
the observations of heavy element ratios in metal-poor stars in dwarf spheroidal galaxies 
\citep{Tsujimoto14} suggested the mass of $\sim 0.01$ M$_\odot$ for NSM ejecta.
The abundance pattern of the r-process isotopes shows the yield ratio of 
the 1st peak r-process isotopes to the heavier ones is about four \citep{Sneden08}.
Therefore, additional $\sim 0.04$ M$_\odot$ 1st peak r-process isotopes from an ultra-stripped SN 
are necessary to produce the solar r-process abundance pattern.
In this study, the yield of the 1st peak r-process isotopes in an ultra-stripped SN is less than 
0.01 M$_\odot$.
If we assume that all binary neutron stars, which merge within the cosmological age, are formed 
after the explosions of ultra-stripped SNe, this yield could explain 25\% of the solar r-process 
abundance at most.
However, we also need other sources for producing the 1st peak r-process isotopes.
We will discuss ECSNe and weak SNe from light Fe cores as other candidates for the sources 
of light r-process isotopes in \S 4.4.

Although NSMs are now the most promising site of heavy r-process elements in very metal-poor stars,
the origin of the 1st peak r-process isotopes is still in debate.
\citet{Tsujimoto14b} proposed that the 1st peak r-process isotopes have been produced 
in core-collapse SNe rather than NSMs from the observed anti-correlation of the abundance 
ratios of Eu/Fe and Sr/Fe.
However, some outlier stars having the enrichment of Sr/Fe compared with the anti-correlation and 
large Eu/Fe ratio were observed.
They considered that these stars have evidence for the enrichment of both light and heavy r-process
isotopes by NSMs.
Here, we showed the possibility that ultra-stripped SNe occurred before NSMs also 
produce the 1st peak r-process isotopes.
Although there is an uncertainty in the abundance distribution of the r-process isotopes,  
ultra-stripped SNe could be a site for light r-process isotopes of such very metal-poor stars.

\subsection{A small CO core as a progenitor of a weak SN evolved from a single massive star}

A star having a CO core slightly heavier than the progenitor of an ECSN ignites Ne at off-center region 
\citep{Woosley80, Nomoto88, Umeda12, Woosley15}.
The evolutionary path of such a star after the C burning is similar to that of the progenitors of 
ultra-stripped SNe. 
Therefore, the SN explosion for it is expected to be weak like an ultra-stripped SN and the SN would 
also produce the 1st peak r-process isotopes.
Here, we discuss the production of the 1st peak r-process isotopes in weak SNe evolved from 
single massive stars having a CO core slightly heavier than an ECSN progenitor.

We discuss the initial mass range of weak SNe having a light CO core assuming that the yield of 
the 1st peak r-process isotopes of a weak SN is $M_{\rm R}$ M$_\odot$.
We roughly estimate the total amount of the 1st peak r-process isotopes in the Galaxy as 
$\sim 2 \times 10^4$ M$_\odot$ from the heavy r-element amount in the Galaxy \citep{Hotokezaka15} 
and the r-process abundance distribution \citep{Sneden08}.
Considering the above and the age of the Universe ($\sim 1.4 \times 10^{10}$ yr), 
we deduce the rate of weak SNe as $ \sim (7 \times 10^{5} M_{\rm R})^{-1}$ yr$^{-1}$.
If the SN rate in the Galaxy is $\sim 0.01$ yr$^{-1}$, the ratio of weak SNe in all SNe is 
$R_{\rm wk} \sim (7 \times 10^3 M_{\rm R})^{-1}$.
In this case, the yield of the 1st peak r-process isotopes is 
$\sim 1.4 \times 10^{-4} R_{\rm wk}^{-1}$ M$_\odot$ 
if all 1st peak r-process isotopes are served by weak SNe.
When we assume the Salpeter initial mass function, $N(M) \propto M^{-2.35}$, the mass range of weak
SN progenitors is in the range between $M_{L}$ and $(1-R_{\rm wk})^{-1/1.35} M_{L}$, 
where $M_L$ is the lowest initial mass of weak SN progenitors.
If the lower limit is 10 M$_\odot$ and the yield of the 1st peak r-process isotopes 
is 0.01, $10^{-3}$, and $5 \times 10^{-4}$ M$_\odot$, the upper limit of the progenitor mass 
of weak SNe is $\sim 10.1$, 11.2, and 12.8 M$_\odot$.
This rough estimate indicates that the initial mass range of weak SNe like ultra-stripped SNe 
could be very narrow, but it also could be one to a few M$_\odot$.

Recently, the advanced evolution of single stars covering the progenitors of ECSNe and core-collapse 
SNe has been investigated and their fates were discussed \citep{Umeda12, Woosley15}.
The initial mass range of the stars that experience the off-center O/Ne ignition is about 1 $M_\odot$
although the lower end of the mass is different in these studies.
Hence, the mass range does not change largely even if one considers uncertainties by different 
convective treatment in stellar evolution models or the effect of stellar rotation.
In this case, 
the production of $10^{-3}$ M$_\odot$ 1st peak r-process isotopes in a weak SN can explain
the origin of the 1st peak r-process isotopes in the Galaxy.

The smaller yield of the 1st peak r-process isotopes from a single weak SN could be caused 
by differences in the progenitor structure.
The main difference between a weak SN from a single star having a light CO core and an ultra-stripped SN
is the possession of the He layer and the H-rich envelope.
Since single stars have outer layers, a part of the wind materials would be decelerated by the outer layers 
and fall-back to the central region.
In this case, the ejected amount would be reduced.
Further, the ejected wind materials would be suffered by stronger neutrino irradiation to heat and eject 
accreting materials.
Strong neutrino irradiation raises the electron fraction of the wind materials and, thus, it prevents 
the production of heavy elements.
This process would also suppress heavy element production.

\section{Conclusions}

We have investigated the explosive nucleosynthesis of ultra-stripped Type Ic SNe of CO145 and CO15
models.
By weak SN explosions of $\sim 10^{50}$ erg, the materials less than 0.1 M$_\odot$ having
shock-heated component and neutrino-irradiated winds are ejected.
An ultra-stripped SNe releases $^{56}$Ni of 0.006--0.01 M$_\odot$ in our models.
These SNe will be observed as fast decaying Type Ic SNe with a peak magnitude of $-15$--$-16$.
The energy release by the radioactive decays of intermediate and heavy elements also 
contribute to the optical emission for several days after the explosion.
The expected light curves are less luminous than a fast decaying Type Ic SN 2005ek.
If the explosion energy is larger and more $^{56}$Ni is produced, ultra-stripped SNe
could be more luminous.
Ultra-stripped SN is a possible candidate for forming a class of binary neutron stars.
Future observations of ultra-stripped Type Ib/Ic SNe will give a constraint of the merger rate for this class 
of binary neutron stars.

Light trans-iron elements are produced mainly in the neutrino-irradiated ejecta of ultra-stripped SNe.
Although the elements of Ga--Zr are produced, their abundance distribution and total yield
are not clearly determined because of the uncertainty in the electron fraction distribution of the
wind component.
The obtained yield of the 1st peak r-process isotopes is less than 0.01 M$_\odot$ but it could be 
much smaller.
The explosion of the SNe evolved from single stars having a light CO core will
also be weak similarly to ultra-stripped SNe. 
Ultra-stripped SNe and weak SNe having a light CO core are possible production sites for 
light r-process elements during the Galactic chemical evolution and in the solar-system.
These SNe could also produce neutron-rich intermediate isotopes $^{48}$Ca.

\section*{Acknowledgements}

We thank Shinya Wanajo, Ko Nakamura, Tomoya Takiwaki, and Hans-Thomas Janka 
for valuable discussions.
TY acknowledges the hospitality of the Yukawa Institute for Theoretical Physics at Kyoto University
during YIPQS long-term and Nishinomiya-Yukawa memorial workshop on ^^ ^^ Nuclear Physics, 
Compact Stars, and Compact Star Mergers 2016" (YITP-T-16-02).
Numerical computations in this study were in part carried out on XC30 at CfCA in NAOJ and 
SR 16000 and XC40 at YITP in Kyoto University.
KT was supported by research fellowships of JSPS for Young Scientists.
This work was supported by Grant-in-Aid for Scientific Research on Innovative Areas (No. 26104007) 
and the grants-in-aid for Scientific Research 
(Nos. 24244028, 26400220, 26400271, 16H02183, 16H00869, 16K17665), 
MEXT and JICFuS as a priority issue (Elucidation of the fundamental laws and evolution of the universe) 
to be tackled by using Post ^^ K' Computer.








\appendix

\section{Electron-capture SN yield}

In order to see systematic differences among explosion models, we perform a two dimensional 
numerical simulation for the explosion of an ECSN using our radiation-hydrodynamics code.
The ECSN progenitor is adopted from a 1.3776 M$_\odot$ O-Ne-Mg core evolved 
from an 8.8 M$_\odot$ star \citep{Nomoto87}.
The simulation method of the SN is the same as the explosion simulation of the ultra-stripped SNe 
in this study.
We pursue the explosion for 667 ms after the core bounce.
We obtained the ejecta mass of $5.4 \times 10^{-2}$ M$_\odot$ and the explosion energy of
$3.2 \times 10^{50}$ erg.

\begin{figure}
\begin{center}
\includegraphics[width=6cm,angle=270]{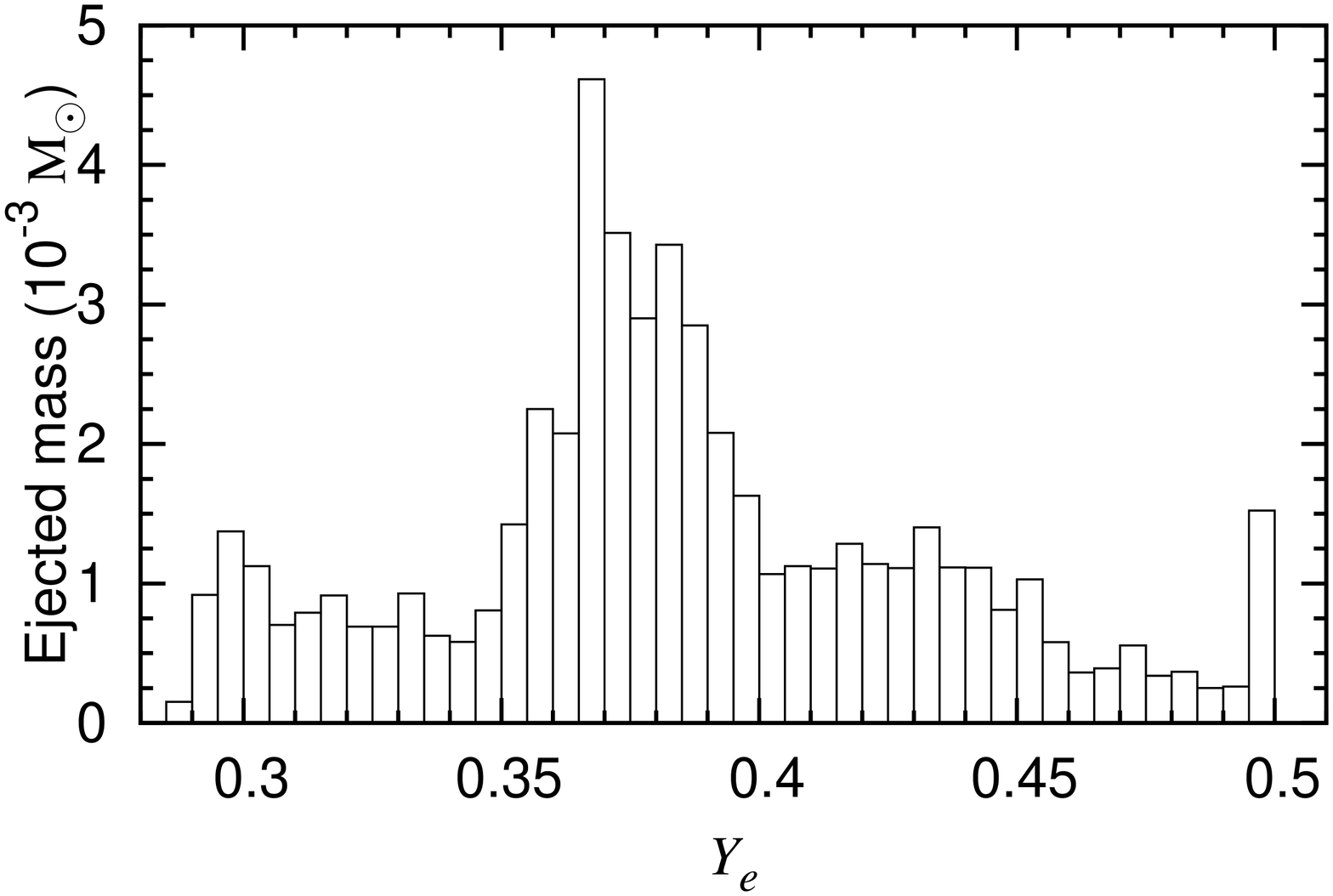}
\end{center}
\caption{
The ejected mass in each bin of $Y_e$ at the initial time of the nucleosynthesis
calculation of the ECSN model.
The vertical axis indicates the ejected mass in units of $10^{-3}$ M$_\odot$ within each $Y_e$ bin 
in the range of $\Delta Y_e = 0.005$.
}
\label{fig:mass-ecsn}
\end{figure}

Then, we calculate the explosive nucleosynthesis of the ECSN ejecta using 13331 tracer fluid particles
having a positive energy and positive radial velocity.
Figure \ref{fig:mass-ecsn} shows the ejected mass in each bin of $Y_e$ 
at the initial time of the nucleosynthesis calculation.
The minimum $Y_e$ value of the ECSN ejecta is 0.288, which is smaller than the corresponding
values found in \citet{Wanajo11} and Janka (2016, private communication; see \S 4.2).
We also see a peak of the ejected mass in the range between $Y_e = 0.35$ and 0.40.
We do not see a peak in the lowest $Y_e$ component.
In the ultra-stripped SN models, the duration between collapse and shock launch is longer than
the ECSN model.
Since the shock launch takes place later, there are some materials staying above PNS 
for a while. Once the shock is launched, the longly staying materials are ejected together with 
materials just captured by the shock.
These two components produce low-$Y_e$ peak in the ultra-stripped SN models.

\begin{figure}
\begin{center}
\includegraphics[width=6cm,angle=270]{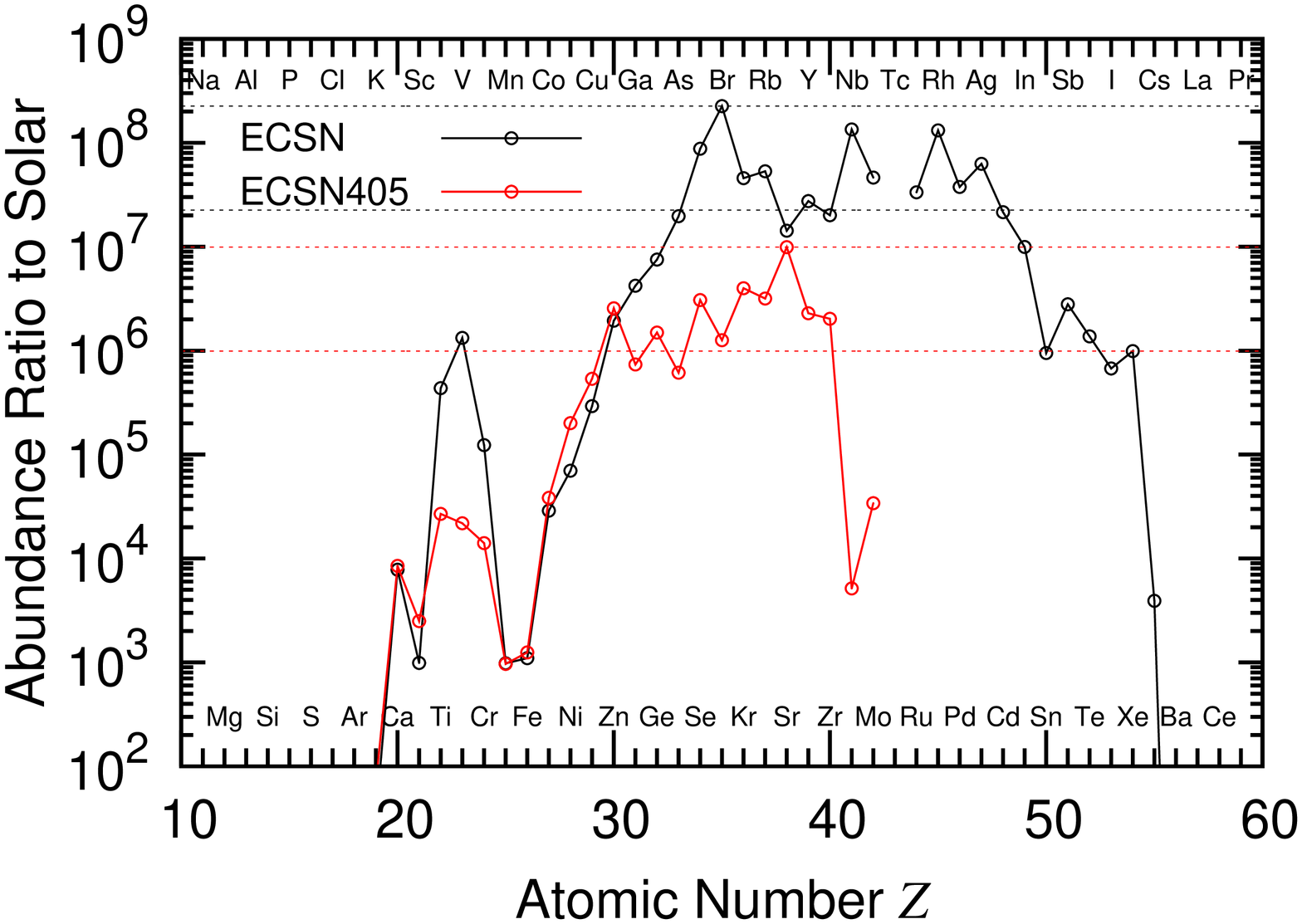}
\end{center}
\caption{
Elemental abundance ratios in the ejecta of the ECSN to the solar abundance.
The black and red lines denote the ratios of the ECSN model and 
the modified $Y_e$ ECSN (ECSN405) model, respectively.
The dashed lines denote the maximum ratios and the ratios of the 10\% of the maximum ratios.
}
\label{fig:ratio-ecsn}
\end{figure}

In our ECSN model, the ejected $^{56}$Ni amount is $1.1 \times 10^{-3}$ M$_\odot$.
This amount is smaller than the corresponding amount found in \citet{Wanajo11} by a factor of three.
The elemental abundance ratio of the ECSN ejecta to the solar abundance is shown in 
Fig. \ref{fig:ratio-ecsn}.
In this model, the elements between As and Cd indicate large abundance ratios.
The ECSN model produces heavier elements than the ultra-stripped SNe.
This is due to the ejection of more neutron-rich wind materials compared to the ejecta of 
the ultra-stripped SNe.
In this calculation, the elements heavier than Mo are produced only in the ejecta of $Y_e \la 0.37$.
The mass of the light trans-iron elements in the ECSN ejecta is $3.3 \times 10^{-2}$ M$_\odot$.
This large yield is also due to the ejection of the neutron-rich materials.
This model also produces heavier elements than the ECSN model in \citet{Wanajo11}.
The abundance distribution of the heavy elements is roughly consistent with their model
with $Y_{e, {\rm min}} = 0.30$.

In order to investigate differences in the abundance distribution by $Y_e$ distribution, we also 
calculate the explosive nucleosynthesis changing the $Y_e$ distribution in accordance with 
Eq. (1) and $Y_{e, {\rm min}} = 0.405$, which is close to the minimum of the ECSN model in 
\citet{Wanajo11}.
We denote this model as ECSN405 model.
This model produces $^{56}$Ni of $1.36 \times 10^{-3}$ M$_\odot$.
The elemental abundance ratio of ECSN405 model to the solar abundance is also shown in
Fig. \ref{fig:ratio-ecsn}.
The light trans-iron elements between Ge and Zr indicate large abundance ratios.
However, the elements heavier than Zr are not produced in this model.
The distribution of the abundance ratio is consistent with the ECSN model in \citet{Wanajo11}.
The total yield of the heavy elements in this model is $3.6 \times 10^{-3}$ M$_\odot$.
This yield is smaller than that of the original model by a factor of nine, so the total yield also 
depends on the $Y_e$ distribution.

In the case of the ECSN, our explosion model indicates larger ejecta mass, larger explosion energy, and
the ejection of more neutron-rich materials compared to the explosion model in \citet{Wanajo11}
\citep[see also][]{Janka08}.
These differences would be due to different treatment on neutrino transport.
The differences found in the SN ejecta would give influences to the explosive nucleosynthesis.
As a result, a larger amount of heavy elements, including the elements heavier than 
light trans-iron elements up to Cd, are produced in our model.
When we change the $Y_e$ distribution to shift the minimum $Y_e$ value to 0.405, the produced 
heavy elements are Ga--Zr.
So, our model seems to be subject to producing neutron-rich materials than the models in 
\citet{Janka08}, which is preferable to produce heavier r-elements.
We should note that the above uncertainties are also included in the ultra-stripped SN models and 
these uncertainties would also affect the abundance distributions of ultra-stripped SNe.

\section{Isotopic abundance ratios in modified electron-fraction cases for CO145 model}

\begin{figure}
\begin{center}
\includegraphics[width=6cm,angle=270]{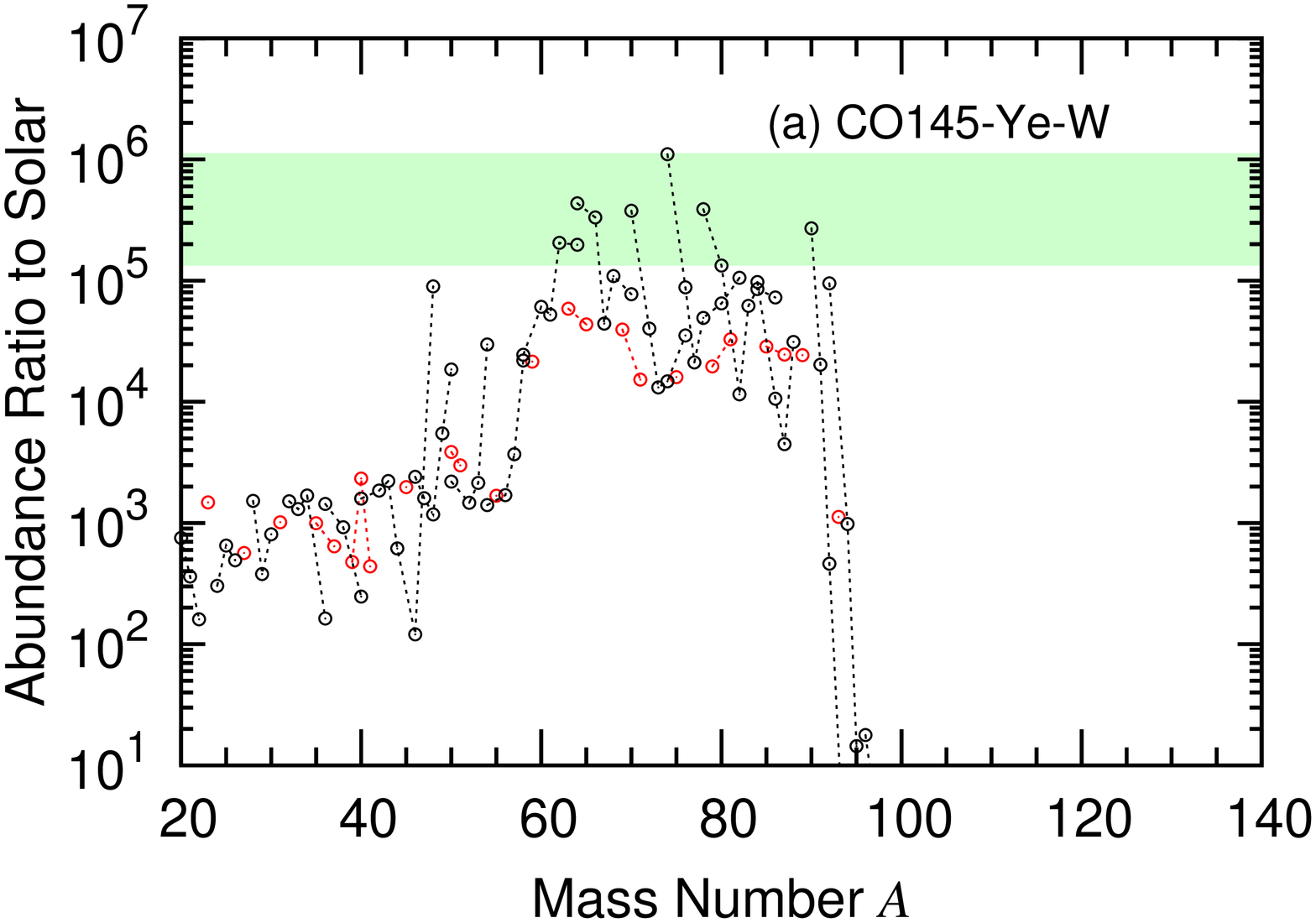}
\includegraphics[width=6cm,angle=270]{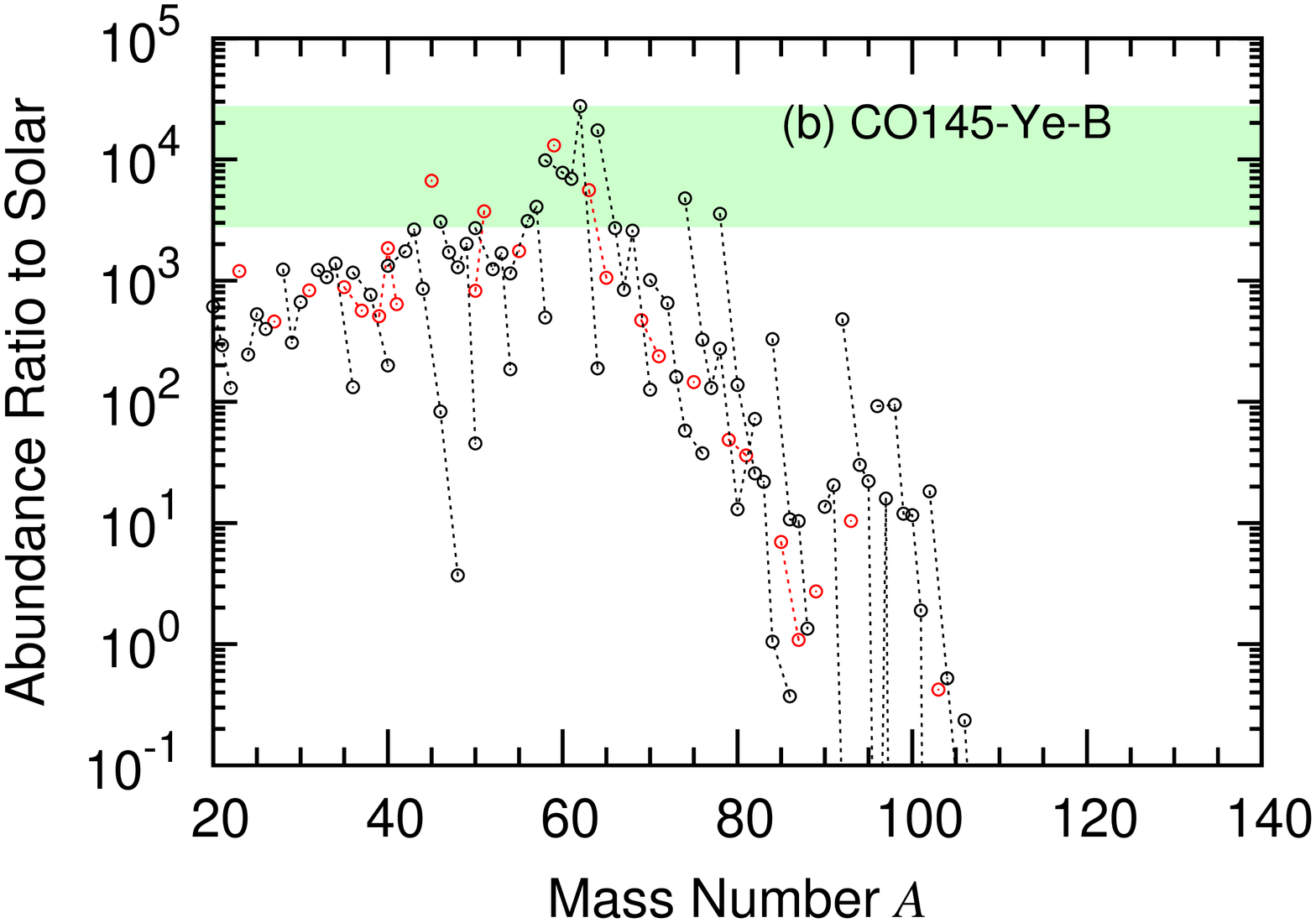}
\end{center}
\caption{
Same as Fig. \ref{fig:abundratio_a}, but for CO145-Ye-W (panel (a)) and CO145-Ye-B (panel (b))
models.
}
\label{fig:abundratio_a-yewb}
\end{figure}

The isotopic abundance ratios to the solar composition in Ye-W and Ye-B models in 
Fig. \ref{fig:abundratio_a-yewb}.
We see in Ye-W model that neutron-rich isotopes are produced in $Z < 30$ elements
and neutron-deficient ones are produced in $Z \ge 30$ elements.
A large abundance ratio is obtained in the neutron-rich isotope $^{48}$Ca.
Heavy-element isotopes $^{90}$Zr and $^{92}$Mo are also produced.
In Ye-B model, neutron-deficient heavy isotopes $^{74}$Se and $^{78}$Kr are produced.
They are mainly produced in the most neutron-rich materials ($Y_e \sim$ 0.47--0.48).


\bsp	
\label{lastpage}
\end{document}